\documentclass[11pt]{article}
\newcommand{\nc}{\newcommand}
\nc{\bib}{\bibitem}
\nc{\al}{\alpha}
\nc{\g}{\gamma}
\nc{\G}{\Gamma}
\nc{\D}{\Delta}
\nc{\eps}{\epsilon}
\nc{\la}{\lambda}
\nc{\La}{\Lambda}
\nc{\var}{\varphi}
\nc{\pa}{\partial}
\nc{\nn}{\nonumber \\ }
\nc{\hf}{\frac{1}{2}}         
\nc{\dz}{\frac{dz}{2\pi i}}
\nc{\bin}[2]{\left (\begin{array}{c} {#1}\\ {#2} \end{array}\right )}
\nc{\ben}{\begin{equation}}
\nc{\een}{\end{equation}}
\nc{\bea}{\begin{eqnarray}}
\nc{\eea}{\end{eqnarray}}
\nc{\bra}[1]{\langle {#1}|}
\nc{\ket}[1]{|{#1}\rangle}
\newcommand{\Z}{\mbox{$Z\hspace{-2mm}Z$}}
\nc{\C}{\mbox{\hspace{1.24mm}\rule{0.2mm}{2.5mm}\hspace{-2.7mm} C}}
\nc{\Nat}{\mbox{\hspace{.04mm}\rule{0.2mm}{2.8mm}\hspace{-1.5mm} N}}
\newcommand{\R}{\mbox{\hspace{.04mm}\rule{0.2mm}{2.8mm}\hspace{-1.5mm} R}}
\def\vvdots{\mathinner{\mkern1mu\raise1pt\vbox{\kern7pt\hbox{.}}\mkern2mu
 \raise4pt\hbox{.}\mkern2mu\raise7pt\hbox{.}\mkern1mu}}

\def\mod{{\rm mod}}
\def\max{{\rm max}}
\def\min{{\rm min}}

\begin{document}

\topmargin -5mm
\oddsidemargin 5mm

\begin{titlepage}
\setcounter{page}{0}

\vspace{8mm}
\begin{center}
{\huge ${\cal N}$-point and higher-genus $osp(1|2)$ fusion}

\vspace{15mm}
{\large J{\o}rgen Rasmussen}\footnote{rasmusse@crm.umontreal.ca}\\[.2cm] 
{\em Centre de recherches math\'ematiques, Universit\'e de Montr\'eal}\\ 
{\em Case postale 6128, 
succursale centre-ville, Montr\'eal, Qu\'ebec, Canada H3C 3J7}

\end{center}

\vspace{10mm}
\centerline{{\bf{Abstract}}}
\vskip.4cm
\noindent
We study affine $osp(1|2)$ fusion, the fusion in $osp(1|2)$ 
conformal field theory, for example.
Higher-point and higher-genus fusion is discussed.
The fusion multiplicities are characterized as discretized volumes
of certain convex polytopes, and are written 
explicitly as multiple sums measuring those volumes.
We extend recent methods developed to treat 
affine $su(2)$ fusion. They are based on the concept of generalized
Berenstein-Zelevinsky triangles and virtual couplings.
Higher-point tensor products of finite-dimensional irreducible $osp(1|2)$
representations are also considered. The associated multiplicities are
computed and written as multiple sums.
\end{titlepage}
\newpage
\renewcommand{\thefootnote}{\arabic{footnote}}
\setcounter{footnote}{0}

\section{Introduction}

The representation theory of finite-dimensional irreducible representations
of the Lie superalgebra $osp(1|2)$ is well-known \cite{PR}.
That includes the decomposition of ordinary three-point tensor products.
However, to the best of our knowledge, the literature does not
offer discussions of general ${\cal N}$-point couplings. Here we shall
consider those and compute the associated tensor product multiplicities.
They are characterized as discretized volumes of certain convex polytopes
(i.e., the number of integer points bounded by the polytope),
and are written explicitly as multiple sums measuring those volumes.
The results are obtained by extending recent methods developed to treat
$su(N)$ tensor products, and are based on the concept of generalized
Berenstein-Zelevinsky (BZ) triangles and virtual couplings 
\cite{virtual,higher}. 

The seminal work by Berenstein and Zelevinsky
\cite{BZ} shows how one may represent ordinary $su(N)$
three-point couplings by triangular arrangements of non-negative integers.
Their results were extended in ref. \cite{higher} to higher-point
couplings, by gluing such triangles together. Here we shall associate two 
types of triangles to $osp(1|2)$
three-point couplings. We denote them $osp(1|2)$ BZ triangles, and 
``super-triangles'', respectively. The gluing method of \cite{higher} is
then applied to treat higher-point $osp(1|2)$ couplings.

We then turn to our second objective: affine $osp(1|2)$ fusion, the fusion
in $osp(1|2)$ conformal field theory, for example. Ordinary three-point
fusion has been studied from various points of view \cite{FY,ERS,osp}.
For integer level, $k$, and associated admissible (or integrable)
representations \cite{KW,FY}, we show that the level dependence
of a fusion may be incorporated in the $osp(1|2)$ BZ triangles or 
super-triangles. That allows us to discuss higher-point fusion along the
lines of \cite{su2} on higher-point $su(2)$ fusion.
As for the tensor products above, the ${\cal N}$-point $osp(1|2)$ 
fusion multiplicities are characterized by level-dependent
convex polytopes, and written explicitly as multiple sums.

Our approach admits also an extension to higher-genus, $h$, fusion.
The associated fusion multiplicities are characterized
as discretized volumes of certain convex polytopes, and
are written explicitly as multiple sums.
To illustrate and demonstrate consistency, we consider in detail the
genus-one one- and two-point fusions.

This work presents the first general results on ${\cal N}$-point $osp(1|2)$ 
tensor products, and on ${\cal N}$-point and higher-genus $osp(1|2)$
fusion. The results are general as they cover all integer 
${\cal N}$, $k$ and $h$.
They are also very explicit and should therefore be easy to use in 
applications. Implementation in computer programming is also straightforward.

\subsection{$osp(1|2)$ representation theory}

Here we recall some basic facts on the Lie superalgebra $osp(1|2)$ and
its irreducible representations \cite{PR}. A ``physicist-friendly'' 
review may be found in ref. \cite{ERSrev}, while general Lie
superalgebra theory is considered in refs. \cite{Kac,FSS}. 

The Lie superalgebra $osp(1|2)$ is a five-dimensional
graded extension of the three-dimensional Lie algebra $su(2)$: 
\bea
 &&\left[J^3,J^\pm\right]=\pm J^\pm,\ \ \ \ 
   \left[J^3,j^\pm\right]=\pm\frac{1}{2}j^\pm,\ \ \ \ 
   \left[J^+,J^-\right]=2J^3\nn
 &&\left\{j^+,j^-\right\}=2J^3,\ \ \ \ \left\{j^\pm,j^\pm\right\}=\pm2J^\pm,
   \ \ \ \ \left[J^\pm,j^\mp\right]=-j^\pm
\label{alg}
\eea
All other (anti-)commutators vanish.
The three even generators $J^+$, $J^-$ and $J^3$ generate an $su(2)$ 
subalgebra of $osp(1|2)$, while $j^+$ and $j^-$ are two odd generators.
They comprise a spin-1/2 representation of the $su(2)$ subalgebra
in the adjoint representation.

Every finite-dimensional irreducible representation has an isospin
$j$ associated to it, where
\ben
 2j\in\Z_{\geq}
\label{2jZ}
\een 
Such a representation ${\cal R}_j$ has dimension $4j+1$:
\ben
 {\cal R}_j:\ \ \ \ \ket{j,j},\ \ket{j,j-1/2}, ...,\ \ket{j,0}, ...,\
  \ket{j,-j+1/2},\ \ket{j,-j}
\label{Rj}
\een
The states $\ket{j,m}$ and $\ket{j,m'}$ have the same parity if and only
if $m-m'\in\Z$. The parity $p({\cal R}_j)$ of the representation
${\cal R}_j$ is defined as the parity of the state $\ket{j,j}$.
The mode $m$ is the eigenvalue of $J^3$: 
$J^3\ket{j,m}=m\ket{j,m}$. It is observed that the representation (\ref{Rj})
splits into two $su(2)$ representations
-- one of spin $j$ and one of spin $j-1/2$. The former consists of the
states $\ket{j,m}$ with $j-m\in\Z_\geq$, while the latter consists of the
states with $j-m\in\Z_\geq+1/2$. Disregarding the notion of parity, the
$osp(1|2)$ representation space (\ref{Rj}) becomes analogous to a single
$su(2)$ representation space of spin $2j$. That observation will turn out to
be useful in the following.

We shall use the same notation $j$ for an $osp(1|2)$ isospin as for an $su(2)$
spin, but refer to them as indicated. An $su(2)$ representation of spin
$j$ is indicated by ${\cal R}_j^{su(2)}$.

\section{Tensor products}

Decompositions of ordinary tensor products of finite-dimensional irreducible 
representations are easily computed:
\ben
 {\cal R}_{j_1}\otimes{\cal R}_{j_2}={\cal R}_{|j_1-j_2|}
  \oplus{\cal R}_{|j_1-j_2|+1/2}\oplus...\oplus{\cal R}_{j_1+j_2-1/2}\oplus
  {\cal R}_{j_1+j_2}
\label{tp}
\een
Note the resemblance to tensor products of integer-spin 
$su(2)$ representations:
\ben
 {\cal R}_{2j_1}^{su(2)}\otimes{\cal R}_{2j_2}^{su(2)}=
  {\cal R}_{2|j_1-j_2|}^{su(2)}
  \oplus{\cal R}_{2|j_1-j_2|+1}^{su(2)}\oplus...\oplus
  {\cal R}_{2(j_1+j_2)-1}^{su(2)}\oplus{\cal R}_{2(j_1+j_2)}^{su(2)}
\label{tpsu2}
\een
Instead of considering a tensor product of the form
\ben
 {\cal R}_{j_1}\otimes{\cal R}_{j_2}\supset{\cal R}_{j_3}
\een
we may equivalently consider the symmetric three-point coupling
to the singlet:
\ben
 {\cal R}_{j_1}\otimes{\cal R}_{j_2}\otimes{\cal R}_{j_3}\supset{\cal R}_{0}
\label{RRR}
\een
Similar couplings of $su(N)$ representations are 
neatly described by Berenstein-Zelevinsky (BZ) triangles \cite{BZ}. 
In the case of $su(2)$ the BZ triangle is trivial but has led to
characterizations of higher-point and higher-genus couplings and
fusions as discretized volumes of certain polytopes \cite{su2}. 
Here we shall discuss the generalization to $osp(1|2)$.

\subsection{Berenstein-Zelevinsky super-triangle}

An $su(2)$ BZ triangle is a triangular arrangement of three non-negative
integer entries $a$, $b$ and $c$
\ben
 {\cal R}_{j_1}^{su(2)}\otimes{\cal R}_{j_2}^{su(2)}\otimes
  {\cal R}_{j_3}^{su(2)}\supset{\cal R}_{0}^{su(2)}
  \ \ \ \ \ \longleftrightarrow\ \ 
\mbox{
\begin{picture}(80,25)(-10,10)
\unitlength=0.5cm
\thicklines
\put(0,0){$c$}
\put(3,0){$a$}
\put(1.5,1.5){$b$}
\end{picture}
}
\label{BZ}
\een\\
subject to the spin constraints
\ben
 a=-j_1+j_2+j_3\in\Z_\geq,\ \ \ \ b=j_1-j_2+j_3\in\Z_\geq,\ \ \ \ 
  c=j_1+j_2-j_3\in\Z_\geq
\label{abc}
\een
and hence 
\ben
 2j_1=b+c, \ \ \ \ 2j_2=c+a,\ \ \ \  2j_3=a+b
\label{abc2}
\een
When all three spins are integer, either $a$, $b$ and $c$
must all be even or all be odd.
Exploring the similarity between (\ref{tp}) and (\ref{tpsu2}) we see
that we may describe three-point couplings of $osp(1|2)$ representations
by standard BZ triangles
\ben
 {\cal R}_{j_1}\otimes{\cal R}_{j_2}\otimes
  {\cal R}_{j_3}\supset{\cal R}_{0}
  \ \ \ \ \ \longleftrightarrow\ \ 
\mbox{
\begin{picture}(80,25)(-10,10)
\unitlength=0.5cm
\thicklines
\put(0,0){$C$}
\put(3,0){$A$}
\put(1.5,1.5){$B$}
\end{picture}
}
\label{BZABC}
\een\\
with isospins
\ben
 4j_1=B+C, \ \ \ \ 4j_2=C+A,\ \ \ \  4j_3=A+B,\ \ \ \ A,\ B,\ C\in\Z_\geq
\label{ABC2}
\een
or equivalently by BZ {\em super-triangles}
\ben
 {\cal R}_{j_1}\otimes{\cal R}_{j_2}\otimes
  {\cal R}_{j_3}\supset{\cal R}_{0}
  \ \ \ \ \ \longleftrightarrow\ \ 
\mbox{
\begin{picture}(80,30)(-10,15)
\unitlength=0.5cm
\thicklines
\put(0,0){c}
\put(4,0){a}
\put(2,2.5){b}
\put(2,1){$\eps$}
\end{picture}
}
\label{sBZ}
\een\\
with isospins
\ben
 2j_1={\rm b}+{\rm c}+\eps,\ \ \ \ 2j_2={\rm c}+{\rm a}+\eps,\ \ \ \ 
  2j_3={\rm a}+{\rm b}+\eps,\ \ \ \ {\rm a,\ b,\ c}\in\Z_\geq,\ \ \ \ 
  \eps\in\left\{0,1\right\}
\label{abceps}
\een
The {\em super-entry} $\eps$ measures the ``parity violation'' of the
coupling:
\bea
 \eps&=&p({\cal R}_{j_1})+p({\cal R}_{j_2})+p({\cal R}_{j_3})
  \ \ \ \mod\ (2)\nn
 &=&2(j_1+j_2+j_3)\ \ \ \mod\ (2)
\label{pv}
\eea

Relaxing the isospin-independent constraints on the entries (thereby allowing
a, b, c, $\eps\in\Z$),
there are infinitely many {\em generalized} super-triangles associated to a
three-point coupling. They are all related through additions of
integer multiples of the (basis) {\em virtual} super-triangle
\ben
 {\cal V}\ =\ \ \ 
\mbox{
\begin{picture}(80,30)(-10,15)
\unitlength=0.5cm
\thicklines
\put(0,0){1}
\put(4,0){1}
\put(2,2.5){1}
\put(2,1){$\bar2$}
\end{picture}
}
\label{V}
\een\\
where $\bar n\equiv-n$. Given an initial generalized super-triangle
${\cal T}_0$ (see (\ref{sD0}) for a choice when extended to higher-point
couplings), all other generalized super-triangles are of the form
\ben
 {\cal T}={\cal T}_0+\sum_{v\in\Z}v{\cal V}
\label{TTV}
\een
However, due to the constraint on $\eps$, only one super-triangle
in this infinite chain of generalized super-triangles will satisfy
all the requirements. We shall call it a {\em true} super-triangle.
By construction, 
if a coupling of three isospins $(j_1,j_2,j_3)$ to the singlet is not 
possible, there will be no true super-triangle associated to that
isospin triplet.

A motivation for introducing super-triangles is that they
seem to indicate how one may generalize the representation of $su(N)$
couplings by BZ triangles to a representation of
Lie superalgebra couplings by (extended)
super-triangles. Even though the $osp(1|2)$ super-triangles are slightly
more complicated to work with than the $osp(1|2)$ BZ triangles, we
shall consider them throughout this paper alongside the BZ triangles.
They provide us with alternative characterizations of tensor product
couplings and fusions -- representations that are more ``supersymmetric''.
Furthermore, in the Conclusion we will indicate how super-triangles appear
natural from the point of view of three-point functions in $osp(1|2)$
conformal field theory.

\subsection{Higher-point couplings}

In a decomposition of a higher-point tensor product, the singlet may occur
more than once, i.e., the associated tensor product multiplicity,
$T_{j_1,...,j_{\cal N}}$, may be greater than one:
\ben
 {\cal R}_{j_1}\otimes...\otimes{\cal R}_{j_{\cal N}}\supset 
  T_{j_1,...,j_{\cal N}}{\cal R}_0
\label{RT}
\een
The similar situation for $su(2)$ couplings is described in \cite{higher,su2}
(ref. \cite{higher} covers all $su(N)$ but does not discuss fusion).
There it is discussed how BZ triangles may be glued together to
form ${\cal N}$-sided diagrams representing the 
$T_{j_1,...,j_{\cal N}}^{su(2)}$ different $su(2)$ couplings.
Likewise, we can associate an ${\cal N}$-sided diagram to each of the
$T_{j_1,...,j_{\cal N}}$ different $osp(1|2)$ ${\cal N}$-point couplings.
Due to the existence of two types of triangles (\ref{BZABC}) and (\ref{sBZ}),
we may represent an $osp(1|2)$ ${\cal N}$-point coupling by two different
types of diagrams. We shall call the ones based on super-triangles
{\em super-diagrams}.

The general method for computing higher-point $su(N)$ tensor product
multiplicities outlined in \cite{higher}, is based on gluing
BZ triangles together using {\em gluing} diagrams (we refer to \cite{higher}
for details). This idea extends readily to $osp(1|2)$ tensor products
(\ref{RT}). To be explicit, let us consider the following 
${\cal N}$-point diagram (in this example ${\cal N}$ is assumed odd):
\ben
\mbox{
\begin{picture}(100,110)(90,-25)
\unitlength=1cm
\thicklines
 \put(0,0){\line(1,2){1}}
 \put(0,0){\line(1,0){2}}
 \put(2,0){\line(-1,2){1}}
\thinlines
\put(1,0.6){\line(0,-1){1}}
\put(1,0.6){\line(2,1){1.6}}
\put(1,0.6){\line(-2,1){1}}
\thicklines
\put(-0.6,1.3){$4j_{\cal N}$}
\put(0.5,-0.9){$4j_{{\cal N}-1}$}
 \put(1.47,2){
\begin{picture}(50,50)
 \put(0,0){\line(1,-2){1}}
 \put(0,0){\line(1,0){2}}
 \put(2,0){\line(-1,-2){1}}
\thinlines
\put(1,-0.6){\line(0,1){1}}
\thicklines
\put(0.5,0.6){$4j_{{\cal N}-2}$}
\end{picture}}
\put(3.2,0){\begin{picture}(50,50)
 \put(0,0){\line(1,2){1}}
 \put(0,0){\line(1,0){2}}
 \put(2,0){\line(-1,2){1}}
\thinlines
\put(1,0.6){\line(0,-1){1}}
\put(1,0.6){\line(2,1){1}}
\put(1,0.6){\line(-2,1){1.6}}
\thicklines
\put(2.67,1.1){$\dots$}
\put(0.5,-0.9){$4j_{{\cal N}-3}$}
\end{picture}}
\put(7,0){\begin{picture}(50,50)
 \put(0,0){\line(1,2){1}}
 \put(0,0){\line(1,0){2}}
 \put(2,0){\line(-1,2){1}}
\thinlines
\put(1,0.6){\line(0,-1){1}}
\put(1,0.6){\line(2,1){1}}
\put(1,0.6){\line(-2,1){1}}
\thicklines
\put(0.8,-0.9){$4j_2$}
\put(2.1,1.2){$4j_1$}
\end{picture}}
\end{picture}
}
\label{jjj}
\een
The role of the gluing is to take care of the summation over internal
isospins in a tractable way. The dual picture of ordinary (Feynman tree-)
graphs is shown in thinner lines. Along a gluing, the opposite isospins
must be identified. 

Let us begin by considering the diagrams obtained by extending (\ref{BZABC}).
The starting point in \cite{higher} and here is
to relax the constraint that the entries 
should be {\em non-negative} integers. As for the super-triangles
above, a diagram of that kind is called a {\em generalized} diagram.
Any such generalized diagram, respecting the gluing constraints and
the outer isospin constraints (\ref{jjj}), will suffice as an initial
diagram. All other diagrams that are associated to the same outer
isospins may then be obtained by adding integer linear combinations
of so-called {\em virtual} diagrams: adding a basis virtual diagram changes
the value of $4j$ of a given internal isospin by two, leaving all other
internal isospins and all outer isospins unchanged. Thus, the basis 
virtual diagram associated to a particular gluing is of the form:
\ben
\mbox{
\begin{picture}(100,65)(-45,-25)
\unitlength=1cm
\thicklines
\put(-2.8,0){${\cal G}\ \ \ =$}
\put(0,0){$\bar1$}
\put(2.4,0){$\bar1$}
 \put(0.8,0.5){$\vvdots$}
\put(1.8,0.9){$\bar1$}
\put(3,0.9){$1$}
\put(-0.6,-0.9){$1$}
\put(0.6,-0.9){$\bar1$}
 \put(1.5,-0.6){$\vvdots$}
\end{picture}
}
\label{gl}
\een
Enumerating the gluing diagrams (\ref{gl}) in (\ref{jjj}) from right to left,
the associated integer coefficients in the
linear combinations are $g_1$,...,$g_{{\cal N}-3}$. If ${\cal D}_0$ 
denotes an initial diagram, all generalized diagrams will then be of the form
\ben
 {\cal D}={\cal D}_0+\sum_{l=1}^{{\cal N}-3}\sum_{g_l\in\Z}g_l{\cal G}_l
\label{DDG}
\een

It remains to be accounted for how to write down an initial diagram 
${\cal D}_0$. However, that is straightforward:
\ben
\mbox{
\begin{picture}(100,130)(100,-25)
\unitlength=1.3cm
\put(-1.5,1){${\cal D}_0\ \ \ =$}
\thicklines
 \put(0,0){\line(1,2){1}}
 \put(0,0){\line(1,0){2}}
 \put(2,0){\line(-1,2){1}}
\thinlines
\put(1,0.6){\line(0,-1){1}}
\put(1,0.6){\line(2,1){1.6}}
\put(1,0.6){\line(-2,1){1}}
\thicklines
\put(-0.5,-0.2){$A$}
\put(0.8,2.3){$B$}
\put(1.7,-0.5){$C$}
 \put(1.47,2){
\begin{picture}(50,50)
 \put(0,0){\line(1,-2){1}}
 \put(0,0){\line(1,0){2}}
 \put(2,0){\line(-1,-2){1}}
\thinlines
\put(1.04,-0.6){\line(0,1){1}}
\thicklines
\put(-0.1,0.3){$4j_{{\cal N}-2}$}
\put(0.6,-2.5){$e_{{\cal N}-3}$}
\end{picture}}
\put(3.2,0){\begin{picture}(50,50)
 \put(0,0){\line(1,2){1}}
 \put(0,0){\line(1,0){2}}
 \put(2,0){\line(-1,2){1}}
\thinlines
\put(1,0.6){\line(0,-1){1}}
\put(1,0.6){\line(2,1){1}}
\put(1,0.6){\line(-2,1){1.6}}
\thicklines
\put(2.67,1.1){$\dots$}
\put(0.2,2.3){$0$}
\put(-0.1,-0.5){$4j_{{\cal N}-3}$}
\put(2,-0.5){$0$}
\put(0.8,2.3){$e_{{\cal N}-4}$}
\end{picture}}
\put(7,0){\begin{picture}(50,50)
 \put(0,0){\line(1,2){1}}
 \put(0,0){\line(1,0){2}}
 \put(2,0){\line(-1,2){1}}
\thinlines
\put(1,0.6){\line(0,-1){1}}
\put(1,0.6){\line(2,1){1}}
\put(1,0.6){\line(-2,1){1}}
\thicklines
\put(-0.1,-0.5){$4j_2$}
\put(1,2.3){$e_1$}
\put(2.3,-0.2){$0$}
\end{picture}}
\end{picture}
}
\label{D0}
\een
with
\bea
 &&e_l=4(j_1+...+j_l),\ \ \ \ 1\leq l\leq{\cal N}-3\nn
 &&A=-S+4j_{{\cal N}-1}+4j_{{\cal N}},\ \ \ \ 
   B=S-4j_{{\cal N}-1},\ \ \ \ 
   C=S-4j_{{\cal N}}
\label{e}
\eea
and
\ben
 S\equiv2(j_1+...+j_{{\cal N}})
\label{S}
\een
Re-imposing the condition that all the entries in ${\cal D}$
(\ref{DDG}) must be {\em non-negative}, results in a set of inequalities
defining a convex polytope in the Euclidean space $\R^{{\cal N}-3}$:
\bea
 0&\leq&g_1,\ 4j_1-g_1,\ 4j_2-g_1\nn
 0&\leq&g_2-g_1,\ 4j_3-g_2+g_1,\ 4(j_1+j_2)-g_2-g_1\nn
 &\vdots&\nn
 0&\leq&g_{{\cal N}-3}-g_{{\cal N}-4},\ 4j_{{\cal N}-2}-g_{{\cal N}-3}
  +g_{{\cal N}-4},\ 4(j_1+...+j_{{\cal N}-3})-g_{{\cal N}-3}
  -g_{{\cal N}-4}\nn
 0&\leq&S-4j_{{\cal N}-1}-g_{{\cal N}-3},\ 
  S-4j_{{\cal N}}-g_{{\cal N}-3},\ -S+4(j_{{\cal N}-1}
   +j_{{\cal N}})+g_{{\cal N}-3}
\label{pol}
\eea
By construction, its discretized volume is the tensor product multiplicity
$T_{j_1,...,j_{\cal N}}$. The volume may be measured explicitly,
expressing the multiplicity as a multiple sum. 
In order to avoid discussing intersection of faces we have to choose an
``appropriate order'' of summation (see refs. \cite{virtual,higher,su2}).
Making such a choice is straightforward, and we find that  
the $osp(1|2)$ tensor product multiplicity $T_{j_1,...,j_{\cal N}}$ may
be written as
\bea
 T_{j_1,...,j_{\cal N}}&=&\sum_{g_{{\cal N}-3}=
   S-4(j_{{\cal N}-1}+j_{{\cal N}})}^{\min\{S-4j_{{\cal N}-1},\ 
   S-4j_{{\cal N}}\}}\ 
  \sum_{g_{{\cal N}-4}=-4j_{{\cal N}-2}+g_{{\cal N}-3}}^{\min\{
  g_{{\cal N}-3},\ 4(j_1+...+j_{{\cal N}-3})-g_{{\cal N}-3}\}}\ \times\ ...\nn
 &&\times\sum_{g_2=-4j_{4}+g_3}^{\min\{g_3,\ 4(j_1+...+j_{3})-g_3\}}\
  \sum_{g_1=\max\{0,\ -4j_{3}+g_2\}}^{\min\{4j_1,\ 4j_{2},
  \ g_2,\ 4(j_1+j_{2})-g_2\}}1
\label{sum}
\eea
This is the first general result for higher-point $osp(1|2)$ tensor
product multiplicities.

Following methods discussed in \cite{virtual,higher,su2}, it is not difficult
to derive necessary and sufficient conditions determining when an
$osp(1|2)$ ${\cal N}$-point tensor product multiplicity is non-vanishing.
The conditions are
\ben
 2j_l,\ S-4j_l\in\Z_{\geq},\ \ \ \ l=1,...,{\cal N}
\label{nonv}
\een
with $S$ defined in (\ref{S}).

Gluing super-triangles together to represent higher-point couplings,
is not a lucrative alternative to the method above. Nevertheless,
we give here the associated gluing super-diagram:
\ben
\mbox{
\begin{picture}(100,90)(-30,-35)
\unitlength=1.5cm
\thicklines
\put(-2,0){${\cal G}\ \ \ =$}
\put(0,0){$0$}
\put(2.4,0){$0$}
 \put(0.8,0.5){$\vvdots$}
\put(1.8,0.9){$0$}
  \put(2.4,0.5){$\bar1$}
\put(3,0.9){$1$}
\put(-0.6,-0.9){$1$}
\put(0.6,-0.9){$0$}
  \put(0,-0.5){$\bar1$}
 \put(1.5,-0.6){$\vvdots$}
\end{picture}
}
\label{sgl}
\een
There is a virtual super-triangle associated to each
of the glued super-triangles, i.e., there are ${\cal N}-2$ (basis) virtual 
super-diagrams associated to an ${\cal N}$-point super-diagram.
In a self-explaining notation we then have that any generalized 
super-diagram may be written
\ben
 {\cal D}={\cal D}_0+\sum_{l=1}^{{\cal N}-2}\sum_{v_l\in\Z}v_l{\cal V}_l
  +\sum_{l=1}^{{\cal N}-3}\sum_{g_l\in\Z}g_l{\cal G}_l
\label{DDVG}
\een

Now, recall that the super-entry measures the 
parity violation as indicated in (\ref{pv}).  
For ${\cal N}$-point couplings it is the sum of the ${\cal N}-2$
super-entries that measures the parity violation. It is therefore
natural to introduce the {\em parity parameter} $\eta$ 
\ben
 2\eta=\left(\sum_{l=1}^{{\cal N}-2}\eps_l\right)
   \ \ \mod\ (2)\ =\left\{\begin{array}{ll}
  0\ \ {\rm for}\ S\in2\Z_\geq\\
  1\ \ {\rm for}\ S\in2\Z_\geq+1  \end{array}\right.
\label{eta}
\een
which of course must depend only on the outer isospins 
(through $S$ (\ref{S})). We may now write down an initial super-diagram:
\ben
\mbox{
\begin{picture}(100,130)(100,-25)
\unitlength=1.3cm
\put(-1.5,1){${\cal D}_0\ \ \ =$}
\thicklines
 \put(0,0){\line(1,2){1}}
 \put(0,0){\line(1,0){2}}
 \put(2,0){\line(-1,2){1}}
\thinlines
\put(1,0.6){\line(0,-1){1}}
\put(1,0.6){\line(2,1){1.6}}
\put(1,0.6){\line(-2,1){1}}
\thicklines
\put(0.85,0.8){$2\eta$}
\put(-0.5,-0.2){a}
\put(0.8,2.3){b}
\put(1.7,-0.5){c}
 \put(1.47,2){
\begin{picture}(50,50)
 \put(0,0){\line(1,-2){1}}
 \put(0,0){\line(1,0){2}}
 \put(2,0){\line(-1,-2){1}}
\thinlines
\put(1.04,-0.6){\line(0,1){1}}
\put(0.95,-1){$0$}
\thicklines
\put(-0.1,0.3){$2j_{{\cal N}-2}$}
\put(0.6,-2.5){e$_{{\cal N}-3}$}
\end{picture}}
\put(3.2,0){\begin{picture}(50,50)
 \put(0,0){\line(1,2){1}}
 \put(0,0){\line(1,0){2}}
 \put(2,0){\line(-1,2){1}}
\thinlines
\put(1,0.6){\line(0,-1){1}}
\put(1,0.6){\line(2,1){1}}
\put(1,0.6){\line(-2,1){1.6}}
\thicklines
\put(0.95,0.8){$0$}
\put(2.67,1.1){$\dots$}
\put(0.2,2.3){$0$}
\put(-0.1,-0.5){$2j_{{\cal N}-3}$}
\put(2,-0.5){$0$}
\put(0.8,2.3){e$_{{\cal N}-4}$}
\end{picture}}
\put(7,0){\begin{picture}(50,50)
 \put(0,0){\line(1,2){1}}
 \put(0,0){\line(1,0){2}}
 \put(2,0){\line(-1,2){1}}
\thinlines
\put(1,0.6){\line(0,-1){1}}
\put(1,0.6){\line(2,1){1}}
\put(1,0.6){\line(-2,1){1}}
\thicklines
\put(0.95,0.8){$0$}
\put(-0.1,-0.5){$2j_2$}
\put(1,2.3){e$_1$}
\put(2.3,-0.2){$0$}
\end{picture}}
\end{picture}
}
\label{sD0}
\een
with
\bea
 &&{\rm e}_l=2(j_1+...+j_l),\ \ \ \ 1\leq l\leq{\cal N}-3\nn
 &&{\rm a}=-\frac{S}{2}-\eta+2(j_{{\cal N}-1}+j_{{\cal N}}),\ \ \ \ 
   {\rm b}=\frac{S}{2}-\eta-2j_{{\cal N}-1},\ \ \ \ 
   {\rm c}=\frac{S}{2}-\eta-2j_{{\cal N}}
\label{se}
\eea
Since $\frac{S}{2}-\eta=[S/2]$ ($[x]$ denotes the integer value of $x$, i.e.,
the greatest integer less than or equal to $x$), 
the entries a, b and c are integers.
Imposing the condition that the diagram (\ref{DDVG}) must be true,
leads to a set of inequalities in the parameters $v$ and $g$
defining a convex polytope as (\ref{pol}). This polytope is 
embedded in the Euclidean space $\R^{2{\cal N}-5}$. 
The inequalities are straightforward to write down, but are not given here.

\subsection{Four-point couplings}

To illustrate the results above we shall compute the $osp(1|2)$ four-point
tensor product multiplicity $T_{j_1,j_2,j_3,j_4}$. We shall do it in two
ways: first by reducing the general result (\ref{pol}) and (\ref{sum})
to ${\cal N}=4$, and then by gluing super-triangles together. 

It follows from (\ref{pol}) that
\ben
 0\leq g,\ 4j_1-g,\ 4j_2-g,\ S-4j_3-g,\ S-4j_4-g,\ -S+4(j_3+j_4)+g
\label{g}
\een
and therefore
\bea
 T_{j_1,j_2,j_3,j_4}&=&\sum_{g=\max\{0,\ 2(j_1+j_2-j_3-j_4)\}}^{\min\{
  2(j_1+j_2+j_3-j_4),\ 2(j_1+j_2-j_3+j_4),\ 4j_1,\ 4j_2\}}1\nn
 &=&1+\min\{4j_1,\ ...,\ 4j_4,\ S-4j_1,\ ...,\ S-4j_4\}
\label{T4}
\eea
provided the conditions (\ref{nonv}) are satisfied.

Now we turn to the super-triangle approach. For ${\cal N}=4$, the convex
polytope defined by (\ref{DDVG}) and (\ref{sD0}) becomes
\bea
 &&0\leq v_1+g,\ 2j_1+v_1,\ 2j_2+v_1\nn
 &&0\leq-g-2v_1\leq1\nn
 &&0\leq-\left[\frac{S+1}{2}\right]+2(j_3+j_4)+v_2+g,\ \left[\frac{S}{2}\right]
  -2j_3+v_2,\ \left[\frac{S}{2}\right]-2j_4+v_2\nn
 &&0\leq2\eta-g-2v_2\leq1
\label{spol}
\eea 
Note that the inequalities $0\leq\eps_1,\eps_2\leq1$ fix $v_1$ and
$v_2$ in terms of $g$:
\ben
 v_1=-\left[\frac{g+1}{2}\right],\ \ \ \ v_2=-\left[\frac{g+1-2\eta}{2}\right]
\label{v1v2}
\een
That means that the set of inequalities in $g$, $v_1$ and $v_2$ reduces 
to a set of inequalities in the gluing coordinate $g$ alone.
It is not hard to verify that the associated (one-dimensional)
polytope is identical to (\ref{g}).
Thus, the two ways of counting the tensor product multiplicity
$T_{j_1,j_2,j_3,j_4}$ are essentially equivalent. That generalizes to 
${\cal N}$-point couplings.

\section{Fusion}

Here we shall extend the above discussion on tensor products to
affine fusion, fusion in $osp(1|2)$ conformal field theory, for
example. To distinguish this consideration from the similar one concerning
tensor products, we denote finite-dimensional irreducible affine modules
of isospin $j$ by $M_j$. The fusion of three such modules to the
singlet is written (cf. the analogous three-point coupling (\ref{RRR}))
\ben
 M_{j_1}\times M_{j_2}\times M_{j_3}\supset N_{j_1,j_2,j_3}^{(k)}
  M_0
\label{MMM}
\een
The fusion multiplicity $N_{j_1,j_2,j_3}^{(k)}$ depends on the
level $k$, where $k$ characterizes the affine extension of $osp(1|2)$
that turns it into a level-$k$ affine Lie superalgebra.
We shall consider only $k$ a positive integer, and the so-called
admissible (or integrable) representations \cite{KW,FY}. 
They are (for $k$ a positive integer) characterized by
\ben
 2j\in\Z_\geq,\ \ \ \ 2j\leq k
\label{adm}
\een 
The ordinary fusion multiplicities are well-known in that case \cite{FY,ERS}:
\ben
 N_{j_1,j_2,j_3}^{(k\geq j_1+j_2+j_3-1/2)}=T_{j_1,j_2,j_3},\ \ \ \ 
 N_{j_1,j_2,j_3}^{(k<j_1+j_2+j_3-1/2)}=0
\label{NT}
\een
We recall that a non-vanishing three-point tensor product multiplicity 
is one. The non-vanishing conditions follow immediately from (\ref{tp}).

The threshold level, $t$, of a three-point coupling is the minimum level 
at which the coupling appears in fusion \cite{CMW}. 
This means, in particular, that $t$ is integer and that $t\leq k$
for the coupling to appear. From (\ref{NT}),
it is straightforward to determine the threshold level 
of an $osp(1|2)$ coupling of three isospins $(j_1,j_2,j_3)$:
\ben
 t=\left[\frac{S}{2}\right]
\label{t}
\een
One may also assign a threshold level 
to an $osp(1|2)$ BZ triangle or super-triangle. It is known how 
to do that for $su(N\leq4)$ \cite{KMSW,BKMW} and has been explored further 
in \cite{su34}. To the BZ $osp(1|2)$ 
triangle (\ref{BZABC}) we may assign the threshold level 
\ben
 t=\left[\frac{A+B+C}{2}\right]
\label{tABC}
\een
and to the super-triangle (\ref{sBZ}) we may assign the threshold level 
\ben
 t={\rm a}+{\rm b}+{\rm c}+\eps
\label{tsBZ}
\een
Since $t$ is integer, the condition $t\leq k$ on (\ref{tABC}) is
equivalent to 
\ben
 A+B+C-1\leq 2k
\label{2k}
\een
A higher-point coupling can also be assigned a threshold level \cite{su2}.
It is defined in the same way as for three-point couplings.

Recently, efforts have been made to characterize fusion multiplicities in 
terms of polytopes. Most results so far pertain to three-point
fusion \cite{su34,BCLM}, but also higher-genus and higher-point $su(2)$ 
fusions have been discussed \cite{su2}.
Below we shall extend the latter results to $osp(1|2)$.

\subsection{Higher-point fusion}

We are now in a position to discuss ${\cal N}$-point fusion.
Using $osp(1|2)$ BZ triangles, we see that fusion is described by 
supplementing the set of inequalities (\ref{pol}) by ${\cal N}-2$
conditions like (\ref{2k}) -- a condition associated to each of the 
${\cal N}-2$ participating triangles. Thus, an ${\cal N}$-point
fusion is characterized by the inequalities
\bea
 0&\leq&g_1,\ 4j_1-g_1,\ 4j_2-g_1,\ 2k-4(j_1+j_2)+g_1+1\nn
 0&\leq&g_2-g_1,\ 4j_3-g_2+g_1,\ 4(j_1+j_2)-g_2-g_1,\ 2k-4(j_1+j_2+j_3)
  +g_1+g_2+1\nn
 &\vdots&\nn
 0&\leq&g_{{\cal N}-3}-g_{{\cal N}-4},\ 4j_{{\cal N}-2}-g_{{\cal N}-3}
  +g_{{\cal N}-4},\ 4(j_1+...+j_{{\cal N}-3})-g_{{\cal N}-3}
  -g_{{\cal N}-4},\nn
 &&2k-4(j_1+...+j_{{\cal N}-2})+g_{{\cal N}-3}+g_{{\cal N}-4}+1\nn
 0&\leq&S-4j_{{\cal N}-1}-g_{{\cal N}-3},\ 
  S-4j_{{\cal N}}-g_{{\cal N}-3},\ -S+4(j_{{\cal N}-1}
   +j_{{\cal N}})+g_{{\cal N}-3},\nn
 &&2k-S+g_{{\cal N}-3}+1
\label{Npol}
\eea
defining a convex polytope embedded in $\R^{{\cal N}-3}$.
Its discretized volume is the fusion multiplicity 
$N_{j_1,...,j_{\cal N}}^{(k)}$. It may be measured explicitly, 
expressing the multiplicity as a multiple sum:
\bea
 N_{j_1,...,j_{\cal N}}^{(k)}&=&\ \ \ \sum_{g_{{\cal N}-3}=\max\{
   S-4(j_{{\cal N}-1}+j_{{\cal N}}),\ -2k+S-1\}}^{\min\{S-4j_{{\cal N}-1},\ 
   S-4j_{{\cal N}}\}}\nn 
 &&\times\sum_{g_{{\cal N}-4}=\max\{-4j_{{\cal N}-2}+g_{{\cal N}-3},\
  -2k+4(j_1+...+j_{{\cal N}-2})-g_{{\cal N}-3}-1\}}^{\min\{
  g_{{\cal N}-3},\ 4(j_1+...+j_{{\cal N}-3})-g_{{\cal N}-3}\}}\nn
 &&\vdots\nn
 &&\times\sum_{g_2=\max\{-4j_{4}+g_3,\ -2k+4(j_1+...+j_4)-g_3-1\}}^{
   \min\{g_3,\ 4(j_1+...+j_{3})-g_3\}}\nn
 &&\times\sum_{g_1=\max\{0,\ -4j_{3}+g_2,\ -2k+4(j_1+j_2+j_3)-g_2-1\}}^{
  \min\{4j_1,\ 4j_{2},
  \ g_2,\ 4(j_1+j_{2})-g_2\}}1
\label{Nsum}
\eea
This is a new result.

\section{Higher-genus fusion}

Here we will discuss the extension of our results above on genus-zero fusion
to generic genus-$h$ fusion. The results here generalize the
similar ones in ref. \cite{su2} on higher-genus $su(2)$ fusion.
$N_{j_1,...,j_{\cal N}}^{(k,h)}$ denotes the genus-$h$
${\cal N}$-point fusion multiplicity. 

A simple extension of (\ref{jjj}) is the following
genus-$h$ ${\cal N}$-point diagram (in this example ${\cal N}$ is assumed
even, while $h$ is arbitrary):
\ben
\mbox{
\begin{picture}(180,80)(105,-20)
\unitlength=0.5cm
\thicklines
 \put(0,0){\line(1,2){1}}
 \put(0,0){\line(1,0){2}}
 \put(2,0){\line(-1,2){1}}
\thinlines
\put(1,0.6){\line(0,-1){1}}
\put(1,0.6){\line(2,1){1.6}}
\put(1,0.6){\line(-2,1){1}}
\thicklines
\put(-1.4,1.2){$4j_{\cal N}$}
\put(0.1,-1.3){$4j_{{\cal N}-1}$}
 \put(1.34,2){
\begin{picture}(50,50)
 \put(0,0){\line(1,-2){1}}
 \put(0,0){\line(1,0){2}}
 \put(2,0){\line(-1,-2){1}}
\thinlines
\put(1,-0.6){\line(0,1){1}}
\thicklines
\put(0.1,1){$4j_{{\cal N}-2}$}
\end{picture}}
\put(3.2,0){\begin{picture}(50,50)
 \put(0,0){\line(1,2){1}}
 \put(0,0){\line(1,0){2}}
 \put(2,0){\line(-1,2){1}}
\thinlines
\put(1,0.6){\line(0,-1){1}}
\put(1,0.6){\line(2,1){1}}
\put(1,0.6){\line(-2,1){1.6}}
\thicklines
\put(2.45,1.1){$\dots$}
\put(0.1,-1.3){$4j_{{\cal N}-3}$}
\end{picture}}
\put(7,0){\begin{picture}(50,50)
 \put(0,0){\line(1,2){1}}
 \put(0,0){\line(1,0){2}}
 \put(2,0){\line(-1,2){1}}
\thinlines
\put(1,0.6){\line(0,-1){1}}
\put(1,0.6){\line(2,1){0.8}}
\put(1,0.6){\line(-2,1){1}}
\thicklines
\put(0.6,-1.3){$4j_1$}
\put(2.2,-0.1){0}
\put(1.2,1.9){0}
\end{picture}}
\put(10.5,1){\begin{picture}(100,80)
\unitlength=0.9cm
%\thicklines
\put(0,0.5){\line(0,-1){1}}
\put(0,0.5){\line(2,-1){1}}
\put(0,-0.5){\line(2,1){1}}
\put(1.5,0){\line(2,1){1}}
\put(1.5,0){\line(2,-1){1}}
\put(2.5,0.5){\line(0,-1){1}}
\thinlines
\put(1.25,0){\circle{1.4}}
\put(0.55,0){\line(-1,0){1.5}}
\put(1.95,0){\line(1,0){1.05}}
\thicklines
 \put(3.5,0){$\dots$}
\put(5,0.5){\line(0,-1){1}}
\put(5,0.5){\line(2,-1){1}}
\put(5,-0.5){\line(2,1){1}}
\put(6.5,0){\line(2,1){1}}
\put(6.5,0){\line(2,-1){1}}
\put(7.5,0.5){\line(0,-1){1}}
\thinlines
\put(6.25,0){\circle{1.4}}
\put(5.55,0){\line(-1,0){1.05}}
\put(6.95,0){\line(1,0){1.05}}
\thicklines
\put(9.15,0){\line(-2,1){1}}
\put(9.15,0){\line(-2,-1){1}}
\put(8.15,0.5){\line(0,-1){1}}
\thinlines
\put(9.15,0){\circle{1.4}}
\put(8,0){\line(1,0){0.45}}
\end{picture}}
\end{picture}
}
\label{gN}
\een
The dual trivalent fusion graph is represented by thinner lines 
and loops. $h$ is the number of such loops or handles. 
The role of the two zeros in (\ref{gN}) will be discussed below.
The number of internal isospins or gluings is ${\cal N}+3(h-1)$, 
while the number of vertices or triangles is ${\cal N}+2(h-1)$.

First we consider the tadpole diagram
\ben
\mbox{
\begin{picture}(100,40)(-25,-20)
\unitlength=0.7cm
\thicklines
\put(1.15,0){\line(-2,1){1}}
\put(1.15,0){\line(-2,-1){1}}
\put(0.15,0.5){\line(0,-1){1}}
\thinlines
\put(1.15,0){\circle{1.4}}
\put(-0.5,0){\line(1,0){0.95}}
\end{picture}
}
\label{tad}
\een
In terms of $osp(1|2)$ BZ triangles the basis diagram associated to it is
\ben
\mbox{
\begin{picture}(100,30)(-25,-10)
\unitlength=0.7cm
\thicklines
\put(0,0.5){0}
\put(0,-0.5){0}
\put(1,0){2}
\end{picture}
}
\label{gl2}
\een
We call (\ref{gl2}) a {\em loop-gluing} diagram.
Since we are gluing over even integers, the initial tadpole diagram
will depend on $2j$ being even (indicated by $p=0$) or odd (indicated
by $p=1$). With $l$ being the coefficient to (\ref{gl2}), the polytope
is defined by
\ben
 0\ \leq\ 2j,\ 2j,\ p+2l,\ 2k-4j-p-2l+1
\een
Thus, the genus-one one-point fusion multiplicity becomes
\ben
 N_j^{(k,1)}\ =\ 
  \sum_{l=\left[\frac{-p+1}{2}\right]}^{\left[\frac{2k-4j-p+1}{2}
  \right]}1\ =\ k-2j+1
\label{N11}
\een
{\em irrespective} of $2j$ being even or odd. That independence is novel
compared to the similar situation for $su(2)$ \cite{su2}.

The basis loop-gluing super-diagram associated to (\ref{tad}) is
\ben
\mbox{
\begin{picture}(100,45)(-25,-20)
\unitlength=1.1cm
\thicklines
\put(0,0.5){0}
\put(0,-0.5){0}
\put(1,0){1}
\put(0.5,0){0}
\end{picture}
}
\label{sgl2}
\een

Let us also describe the basis loop-gluing diagrams associated to 
the genus-one two-point fusion
\ben
\mbox{
\begin{picture}(100,50)(0,-20)
\unitlength=1cm
\thicklines
\put(0,0.5){\line(0,-1){1}}
\put(0,0.5){\line(2,-1){1}}
\put(0,-0.5){\line(2,1){1}}
\put(1.5,0){\line(2,1){1}}
\put(1.5,0){\line(2,-1){1}}
\put(2.5,0.5){\line(0,-1){1}}
\thinlines
\put(1.25,0){\circle{1.4}}
\put(0.55,0){\line(-1,0){1.05}}
\put(1.95,0){\line(1,0){1.05}}
\thicklines
%
%\put(-0.9,-0.1){$\la$}
%\put(3.2,-0.1){$\mu$}
\end{picture}
}
\label{two}
\een
In terms of $osp(1|2)$ BZ diagrams there are two basis loop-gluings 
associated to this fusion. They may be represented by the diagrams
\ben
\mbox{
\begin{picture}(100,30)(60,-10)
\unitlength=0.7cm
\thicklines
\put(-2,0){${\cal L}\ =$}
\put(0,0.5){$\bar1$}
\put(0,-0.5){1}
\put(1,0){1}
\put(2,0){1}
\put(3,0.5){$\bar1$}
\put(3,-0.5){1}
\put(6,0){${\cal L}'\ =$}
\put(8,0.5){1}
\put(8,-0.5){$\bar1$}
\put(9,0){1}
\put(10,0){1}
\put(11,0.5){1}
\put(11,-0.5){$\bar1$}
\end{picture}
}
\label{lg}
\een
They differ significantly from the $su(2)$ basis loop-gluing diagrams 
\cite{su2}, as they do not constitute a basis of $su(2)$ loop-gluing 
diagrams. Similarly, the two loop-gluing super-diagrams are
\ben
\mbox{
\begin{picture}(100,50)(90,-20)
\unitlength=1.1cm
\thicklines
\put(-1.5,0){${\cal L}\ =$}
\put(0,0.5){$\bar1$}
\put(0,-0.5){0}
\put(0.5,0){1}
\put(1,0){0}
\put(2,0){0}
\put(2.5,0){1}
\put(3,0.5){$\bar1$}
\put(3,-0.5){0}
\put(5,0){${\cal L}'\ =$}
\put(6.5,0.5){0}
\put(6.5,-0.5){$\bar1$}
\put(7,0){1}
\put(7.5,0){0}
\put(8.5,0){0}
\put(9,0){1}
\put(9.5,0.5){0}
\put(9.5,-0.5){$\bar1$}
\end{picture}
}
\label{slg}
\een

It is noted 
that the choice of loop-gluing basis (\ref{lg}) is not a convenient
one. Had we only been interested in the polytope characterization
of the fusion multiplicity and not an explicit measure of its 
discretized volume, this symmetric basis would suffice. But in order
to be able to choose an appropriate order of summation (i.e., avoid
discussing intersection of faces), we can not allow both diagrams
to affect all the entries of the two triangles. A good but less
symmetric basis is
\ben
\mbox{
\begin{picture}(100,30)(60,-10)
\unitlength=0.7cm
\thicklines
\put(-2,0){${\cal L}\ =$}
\put(0,0.5){$\bar1$}
\put(0,-0.5){1}
\put(1,0){1}
\put(2,0){1}
\put(3,0.5){$\bar1$}
\put(3,-0.5){1}
\put(6,0){${\cal L}^+\ =$}
\put(8,0.5){0}
\put(8,-0.5){0}
\put(9,0){2}
\put(10,0){2}
\put(11,0.5){0}
\put(11,-0.5){0}
\end{picture}
}
\label{lgplus}
\een
where ${\cal L}^+={\cal L}+{\cal L}'$.

As a non-trivial check of our procedure, we now consider the
genus-one two-point fusion in detail using the two different channels
\ben
\mbox{
\begin{picture}(180,60)(40,-30)
\unitlength=1cm
\thicklines
\put(0,0.5){\line(0,-1){1}}
\put(0,0.5){\line(2,-1){1}}
\put(0,-0.5){\line(2,1){1}}
\put(1.5,0){\line(2,1){1}}
\put(1.5,0){\line(2,-1){1}}
\put(2.5,0.5){\line(0,-1){1}}
\thinlines
\put(1.25,0){\circle{1.4}}
\put(0.55,0){\line(-1,0){1.05}}
\put(1.95,0){\line(1,0){1.05}}
\thicklines
\put(-1.2,-0.1){$4j_2$}
\put(3.2,-0.1){$4j_1$}
\put(8,0){\begin{picture}(100,50)
\thicklines
\put(1.15,0){\line(-2,1){1}}
\put(1.15,0){\line(-2,-1){1}}
\put(0.15,0.5){\line(0,-1){1}}
\put(-0.4,0.5){\line(-2,-1){1}}
\put(-0.4,-0.5){\line(-2,1){1}}
\put(-0.4,0.5){\line(0,-1){1}}
\put(-1.45,0.85){$4j_2$}
\put(-1.45,-1){$4j_1$}
\thinlines
\put(1.15,0){\circle{1.4}}
\put(-0.75,0){\line(1,0){1.2}}
\put(-0.75,0){\line(-1,2){0.3}}
\put(-0.75,0){\line(-1,-2){0.3}}
\end{picture}
}
\end{picture}
}
\label{two2}
\een
Consistency requires the associated fusion multiplicities to coincide.
To ensure that we are gluing over even integers, an initial diagram 
associated to the channel on the left depends on
$2j_1+2j_2$ being even (indicated by $p=0$) or odd (indicated by $p=1$).
Using the loop-gluing diagrams (\ref{lgplus}), and writing the inequalities
associated to the rightmost triangle first, we find the polytope
defined by
\bea
 0&\leq&2j_1-l,\ 2j_1+l,\ 2j_2+l+2l^++p,\ 2k-4j_1-2j_2-l-2l^+-p+1\nn
 0&\leq&2j_2-l,\ 2j_2+l,\ 2j_1+l+2l^++p,\ 2k-2j_1-4j_2-l-2l^+-p+1 
\label{twoin}
\eea
It follows that the genus-one two-point fusion multiplicity is
in fact independent of $p$, and is given by:
\ben
 N_{j_1,j_2}^{(k,1)}\ =\ (k-2\max\{j_1,\ j_2\}+1)(4\min\{j_1,\ j_2\}+1)
\label{N12}
\een
It is straightforward to choose an initial diagram associated to the channel
on the right (\ref{two2}) that is independent of $p$. Using the gluing
diagram (\ref{gl}) and the loop-gluing diagram (\ref{gl2}), we are led
to consider the polytope defined by
\bea
 0&\leq&2l+g,\ -g,\ -g,\ 2k-2l+g+1\nn
 0&\leq&2j_1+2j_2+g,\ 2j_1-2j_2-g,\ -2j_1+2j_2-g,\ 2k-2j_1-2j_2+g+1
\label{twoin2}
\eea
Its discretized volume is seen to be (\ref{N12}), as desired.
This result resembles the similar one for $su(2)$ \cite{su2}, but differs 
by involving the two {\em different} factors 2 and 4.

Note that (\ref{N12}) reduces correctly
to (\ref{N11}) for $\min\{j_1,\ j_2\}=0$.
In fact, it is a general feature of fusion that the ${\cal N}$-point fusion
multiplicity $N_{j_1,...,j_{\cal N}}^{(k,h)}$ is equal to the 
$({\cal N}+1)$-point fusion multiplicity $N_{j_1,...,j_{\cal N},0}^{(k,h)}$
(it is further recalled that a fusion multiplicity is symmetric under 
permutations of its lower indices).
That is not an obvious property of our construction, but will be
used below. There we shall restrict to ${\cal N}\geq3$ which accordingly
is not a real restriction. The rationale for doing it, though, is that it
allows us to make a universal choice of initial diagram associated to
the fusion (\ref{gN}). On the other hand, in the case of zero-, one- 
or two-point fusion it results in unnecessarily complicated
polytopes and multiple sums. For the benefit of the
presentation here, we are not including other specific results than
(\ref{N11}) and (\ref{N12}) for
such lower-point fusions. However, they are easily obtained
following our general procedure.

\subsection{General result}

It is now straightforward to write down the inequalities defining 
the convex polytope associated to (\ref{gN}).
Here we focus on the $osp(1|2)$ BZ triangle approach using (\ref{gl}) and
(\ref{lgplus}), in particular.
Our choice of initial diagram is indicated in (\ref{gN}) by the two zeros:
all entries of the higher-genus part to the right of them are zero,
while the ${\cal N}$-point part follows the pattern of the initial diagram
(\ref{D0}) (assuming ${\cal N}\geq3$, see the comments above). 
Enumerating the gluings from right to left (and ${\cal L}^+$ 
before ${\cal L}$), the integer coefficients in the
linear combinations are $-g_1,...,-g_h,g_{h+1},...,g_{{\cal N}+h-2}$
(the sign convention is merely for convenience), and $l_1^+,l_1,...,
l_{h-1}^+,l_{h-1}$, while $l$ is associated to the
tadpole at the extreme right. 
Listing the inequalities associated to the triangles from right to left,
we have the following convex polytope (assuming $h\geq1$):
\bea
 0&\leq&g_1,\ g_1,\ 2l-g_1,\ 2k-g_1-2l+1\nn
 0&\leq&g_1-l_1,\ g_1+l_1,\ -g_1+2l_1^++l_1,\ 2k-g_1-2l_1^+-l_1+1\nn
 0&\leq&g_2-l_1,\ g_2+l_1,\ -g_2+2l_1^++l_1,\ 2k-g_2-2l_1^+-l_1+1\nn
 &\vdots&\nn
 0&\leq&g_{h-1}-l_{h-1},\ g_{h-1}+l_{h-1},\ -g_{h-1}+2l_{h-1}^++l_{h-1},\ 
  2k-g_{h-1}-2l_{h-1}^+-l_{h-1}+1\nn
 0&\leq&g_{h}-l_{h-1},\ g_{h}+l_{h-1},\ -g_{h}+2l_{h-1}^++l_{h-1},\ 
  2k-g_{h}-2l_{h-1}^+-l_{h-1}+1\nn
 0&\leq&g_{h+1}+g_{h},\ -g_{h+1}+g_{h},\ 4j_1-g_{h+1}-g_{h},\ 
  2k-4j_1+g_{h+1}-g_{h}+1\nn
 0&\leq&g_{h+2}-g_{h+1},\ 4j_1-g_{h+2}-g_{h+1},\ 4j_2
   -g_{h+2}+g_{h+1},\nn
  &&2k-4(j_1+j_2)+g_{h+2}+g_{h+1}+1\nn
 &\vdots&\nn
 0&\leq&g_{{\cal N}+h-2}-g_{{\cal N}+h-3},\ 4(j_1+...
   +j_{{\cal N}-3})-g_{{\cal N}+h-2}-g_{{\cal N}+h-3},\nn 
 &&4j_{{\cal N}-2}-g_{{\cal N}+h-2}+g_{{\cal N}+h-3},\ 
  2k-4(j_1+...+j_{{\cal N}-2})+g_{{\cal N}+h-2}
    +g_{{\cal N}+h-3}+1\nn
 0&\leq&S-4j_{{\cal N}-1}-g_{{\cal N}+h-2},\ S-4j_{{\cal N}}
  -g_{{\cal N}+h-2},\ -S+4(j_{{\cal N}-1}+j_{{\cal N}})
   +g_{{\cal N}+h-2},\nn
  &&2k-S+g_{{\cal N}+h-2}+1
\label{polgN}
\eea
By construction, its discretized volume is the fusion multiplicity
$N_{j_1,...,j_{\cal N}}^{(k,h)}$, which then provides the first
characterization of general $osp(1|2)$ fusion multiplicities.
The volume may be measured explicitly expressing 
$N_{j_1,...,j_{\cal N}}^{(k,h)}$ as a multiple sum:
\bea
 N_{j_1,...,j_{\cal N}}^{(k,h)}&=&\sum_{g_{{\cal N}+h-2}}...
  \sum_{g_h}\left(\sum_{l_{h-1}}\sum_{l_{h-1}^+}\sum_{g_{h-1}}\right)...
  \left(\sum_{l_1}\sum_{l_1^+}\sum_{g_1}\right)\sum_l1\ .
\label{sumNh}
\eea
The integer summation variables are bounded according to
\bea
 \left[\frac{g_1+1}{2}\right]\leq&l&\leq \left[\frac{2k-g_1+1}{2}\right]\nn
 |l_1|\leq &g_1&\leq \min\{2l_1^++l_1,\ 2k-2l_1^+-l_1+1\}\nn
 \left[\frac{g_2-l_1+1}{2}\right]\leq &l_1^+&\leq\left[
  \frac{2k-g_2-l_1+1}{2}\right]\nn
 -g_2\leq &l_1&\leq g_2\nn
 &\vdots&\nn
 |l_{h-1}|\leq &g_{h-1}&\leq\min\{2l_{h-1}^++l_{h-1},\nn
 &&\hspace{2cm}2k-2l_{h-1}^+
  -l_{h-1}+1\}\nn
 \left[\frac{g_h-l_{h-1}+1}{2}\right]\leq &l_{h-1}^+&\leq\left[
  \frac{2k-g_h-l_{h-1}+1}{2}\right]\nn
 -g_h\leq &l_{h-1}&\leq g_h\nn
 |g_{h+1}|\leq&g_h&\leq\min\{4j_1-g_{h+1},\ 2k-4j_1+g_{h+1}+1\}\nn
 \max\{-4j_2+g_{h+2},\ \ \ \ \ \hspace{1cm}&&\nn
   -2k+4(j_1+j_2)-g_{h+2}-1\}
  \leq&g_{h+1}&\leq \min\{g_{h+2},\ 4j_1-g_{h+2}\}\nn
 &\vdots&\nn
 \max\{-4j_{{\cal N}-2}+g_{{\cal N}+h-2},\ \ \ \hspace{2cm} &&\nn
  -2k+4(j_1+...+ 
  j_{{\cal N}-2})-g_{{\cal N}+h-2}-1\}\leq &g_{{\cal N}+h-3}&\leq
  \min\{g_{{\cal N}+h-2},\nn
    &&\ \ \hspace{1cm}4(j_1+...+j_{{\cal N}-3})-g_{{\cal N}+h-2}\}\nn
 \max\{S-4(j_{{\cal N}-1}+j_{{\cal N}}),\ -2k+S-1\}\leq
  &g_{{\cal N}+h-2}&\leq\min\{S-4j_{{\cal N}-1},\ S-4j_{{\cal N}}\}
\label{sumNhb}
\eea
This constitutes the first explicit result for the general genus-$h$ 
${\cal N}$-point fusion multiplicities. 

An advantage of using super-triangles instead of the $osp(1|2)$ BZ triangles
employed above, is that the variables
$v$, $g$ and $l$ all appear with unit coefficients in the 
polytope-defining inequalities similar to (\ref{polgN}).
However, it is not straightforward to measure the discretized volume
of that polytope. The reason is similar to the one excluding
the basis (\ref{lg}) as a ``good basis''.

\section{Conclusion}

We have considered higher-point couplings of finite-dimensional
irreducible representations of $osp(1|2)$. The associated tensor
product multiplicities were characterized as discretized volumes
of certain convex polytopes, and written explicitly as multiple sums.
The results are general.

We have also considered affine $osp(1|2)$ fusion. By extending the
results on tensor products, we characterized a general genus-$h$ 
${\cal N}$-point fusion multiplicity as a discretized volume of a 
certain convex polytope, and wrote down an explicit multiple sum 
measuring that volume. That result is also general.

It has been demonstrated, though not emphasized explicitly, 
that a fusion polytope may be embedded in the 
associated tensor product polytope. The reason is that the set of defining 
inequalities of a fusion polytope is obtained by supplementing
the set of defining inequalities of the associated tensor product 
polytope by level-dependent inequalities. That offers a geometric 
interpretation of affine fusion being a truncated tensor product.

In the derivation of our results we have described three-point couplings 
by triangular arrangements of non-negative integers similar to the
$su(2)$ BZ triangles. We introduced two types.  
We based most of our results on a direct adaption of the ordinary 
$su(2)$ BZ triangle. However, we also introduced a super-triangle
and discussed some of its alternative features.
Here we will indicate how it appears natural from the point
of view of correlators in $osp(1|2)$ conformal field theory.
Three-point functions in conformal field theory with affine Lie
group symmetry have been considered in ref. \cite{3point}.
Their level-dependence was subsequently addressed in ref. \cite{level}.
The idea is to associate so-called
elementary polynomials to the elementary couplings appearing in
an expansion of a three-point coupling.
The three-point functions are then constructed as (linear combinations of)
products of those polynomials. The algebraic relations (syzygies) among
the elementary couplings complicate the construction.
In some cases they may be taken into account at the level
of BZ triangles by forbidding certain configurations. In terms
of polynomials that amounts to forbidding certain products,
as there is a correspondence between BZ triangles and polynomials.
As we will show elsewhere \cite{osp}, the situation for $osp(1|2)$ is 
most easily handled using our super-triangles. The constraint on
the super-entry $\eps$ (\ref{abceps}) is neatly encoded by associating
a Grassmann odd polynomial to a super-triangle with $\eps=1$.
This also introduces a natural way of implementing the
$osp(1|2)$ syzygy \cite{SVdJH,BCM}.

\vskip.5cm
\noindent{\em Acknowledgements}
\vskip.1cm
\noindent The author thanks Chris Cummins and Pierre Mathieu for helpful 
comments.

\end{document}